\documentstyle[procl]{article}

\def\Journal#1#2#3#4{{#1} {\bf #2}, #3 (#4)}

\def\NPB{{\em Nucl. Phys.} B}

\def\PLB{{\em Phys. Lett.}  B}

\def\PRL{\em Phys. Rev. Lett.}

\def\PRD{{\em Phys. Rev.} D}

\def\ZPC{{\em Z. Phys.} C}

\def\be{\begin{equation}}

\def\ee{\end{equation}}

\def\bea{\begin{eqnarray}}

\def\eea{\end{eqnarray}}

\def\beq{\begin{equation}}
\def\eeq{\end{equation}}
\def\bea{\begin{eqnarray}}
\def\eea{\end{eqnarray}}
\def\bem{\begin{math}}
\def\eem{\end{math}}
\def\bit{\begin{itemize}}
\def\eit{\end{itemize}}
\def\bla{\begin{flushright}}
\def\ela{\end{flushright}}
\def\qq2{$Q^2$}               % Q2
\def\aa1{$A_1(x,Q^2)$}        % A1
\def\ff1{$F_1(x,Q^2)$}        % F1
\def\gg1{$g_1(x,Q^2)$}        % g1
\begin{document}

\title{ ON THE $Q^2$ DEPENDENCE OF ASYMMETRY $A_1$ }

\author{ A.V. KOTIKOV }

\address{Laboratoire de Physique Theorique ENSLAPP\\ LAPP, B.P. 100,
F-74941, Annecy-le-Vieux Cedex, France\\ and \\
Particle Physics Laboratory, JINR, Dubna, Russia}

\author{ D.V. PESHEKHONOV }

\address{Particle Physics Laboratory, JINR, Dubna, Russia}

%%%%%%%%%%%%%%%%%%%%%%%%%%%%%%%%%%%%%%%%%%%%%%%%%%%%%%%%%%%%%%


%%%%%%%%%%%%%%%%%%%%%%%%%%%%%%%%%%%%%%%%%%%%%%%%%%%%%%%%%%%%%%

\maketitle\abstracts{We analyse the proton and deutron data on spin dependent asymmetry
~\aa1 supposing the DIS structure functions $g_1(x,Q^2)$
and $F_3(x,Q^2)$ have the similar $Q^2$-dependence.
As a result, we have obtained $\Gamma_1^p - \Gamma_1^n = 0.192$ at
$Q^2= 10~{\rm GeV}^2$ and $\Gamma_1^p - \Gamma_1^n = 0.165$ at
$Q^2= 3~{\rm GeV}^2$, in the best agreement with the Bjorken
sum rule predictions.}

%\section{Guidelines}

%\subsection{Producing the Hard Copy}\label{subsec:prod}

An experimental study of the nucleon spin structure is realized by
measuring of the asymmetry $A_1(x,Q^2) = g_1(x,Q^2) / F_1(x,Q^2)$.
The most known theoretical predictions on spin dependent structure
function $g_1(x,Q^2)$ of the nucleon were done by Bjorken \cite{Bj} and
Ellis and Jaffe \cite{EJ} for the so called {\it first moment value}
$\Gamma_1 = \int_0^1 g_1(x) dx$.\\
The calculation of the $\Gamma_1$ value requires the knowledge of
structure function $g_1$ at the same $Q^2$ in the hole $x$ range.
Experimentally asymmetry $A_1$ is measuring at different values of $Q^2$
for different $x$ bins.
An accuracy of the
modern experiments
\cite{EG}
allows to analyze data in the assumption \cite{EK93}
that asymmetry \aa1 is \qq2
independent (structure functions $g_1$ and $F_1$ have the same $Q^2$
dependence). But the tune checking of the Bjorken and Ellis - Jaffe sum
rules requires considering the $Q^2$ dependence of $A_1$ or $g_1$
(for recent studies of the $Q^2$ dependence of $A_1$  see the references
of \cite{KoPe}).\\
This article is based on our observation
that the $Q^2$ dependence of $g_1$
and the spin average structure function $F_3$ is the similar in
a wide $x$ range:
$10^{-2} < x < 1$. At small $x$ it seems that may be not true.\\
To demonstrate the validity of the observation, we note that
the r.h.s. of DGLAP equations for
NS parts of $g_1$ and $F_3$ is the same
(at least in first two orders of the perturbative QCD)
and differs from $F_1$ already in the first subleading order.
For the singlet part of $g_1$ and for $F_3$ the difference between
perturbatively calculated spliting functions
is also negligible
(see \cite{KoPe}).
This observation allows us to conclude the function :
\bea
A_1^*(x) = {g_1(x,Q^2) \over F_3(x,Q^2)} \nonumber
\eea
should be practically $Q^2$ independent at $x>0.01$ and
the asymmetry $A_1$ at $Q^2=<Q^2>$ can be defined than as :
\bea
A_1(x_i,<Q^2>) =  {F_3(x_i,<Q^2>) \over F_3(x_i,Q^2_i)} \cdot
{F_1(x_i,Q^2_i) \over F_1(x_i,<Q^2>)} \cdot A_1(x_i,Q^2_i),
\label{5}
\eea
where $x_i$ ($Q^2_i$) means an experimentally measured value of $x$ ($Q^2$).\\
We use SMC and E143 proton and deuteron data for asymmetry
$A_1(x,Q^2)$\cite{EG}.
To get $F_1(x,Q^2)$ we take NMC parametrization for $F_2(x,Q^2)$.
To get the values of
$F_3(x,Q^2)$ we parametrize the CCFR data
as a function of $x$ and $Q^2$. \\
Using eq.(\ref{5}), we recalculate the SMC
and E143
measured asymmetry of the
proton and deuteron at $Q^2= 10~ {\rm GeV}^2$ and $3~ {\rm GeV}^2$,
which are average $Q^2$ of these experiments respectively
and get the value of $\int g_1(x) dx$ through the
measured $x$ ranges.  To obtain the first moment values
$\Gamma_1^{p(d)}$ we have used an original estimations of SMC and E143
for unmeasured regions.
As the last step we calculate the difference $\Gamma_1^p - \Gamma_1^n =
2 \Gamma_1^p -
2 \Gamma_1^d / (1-1.5 \cdot \omega_D)$ where
$\omega_D=0.05$.
%At $Q^2=10~ (3)~ {\rm GeV}^2$
We get the following
results for $\Gamma_1^p - \Gamma_1^n $:
\bea  &Q^2=10 {\rm GeV}^2 &Q^2=3{\rm GeV}^2 \nonumber \\
&0.199 \pm 0.038~~&0.163 \pm 0.026
~~~~~~~~~~~
\mbox{(SMC \mbox{ and } E143)} \nonumber \\
&0.192~~~~~~~~~~~~&0.165~~~~~~~~~~~ ~~~~~~~~~~~\mbox{(our result)}
\nonumber \\
&0.187 \pm 0.003~~&0.171 \pm 0.008
~~~~~~~~~~~~ \mbox{(Theory)}
\nonumber \eea
%\\
As a conclusion, we would like to note that
our observation that function $A_1^*(x)$ is $Q^2$ independent at large
and intermediate $x$ is supported by good agreement (see
present analysis with other estimations
of the $Q^2$ dependence of the $A_1$.

This work is supported partially by the Russian Fund for Fundamental
Research, Grant N 95-02-04314-a.
%%                 REFERENCES %

\end{document}